\documentstyle[12pt]{article}
\input amssym.def
\input amssym
\def \dst {\displaystyle}

\newcommand{\la} {\lambda}
\newcommand{\al} {\alpha}

\newcommand{\ga} {\gamma}
\newcommand{\bga}{\begin{array}{l}}
\newcommand{\ena}{\end{array}}
\newcommand{\bge}{\begin{equation}}
\newcommand{\ene}{\end{equation}}
\title{ Ricci flat metrics in various dimensions, depending from 2
light-cone parameters, and the Lagrangian for the 2 dimensional reduction of
gravity}
\author{ M. Zyskin\thanks{Courant Institute,  251 Mercer St., Greenwich
Village, New York,  NY 10012; zyskin@physics.rutgers.edu}}
\date{}
\begin{document}
\maketitle

\begin{abstract}
\vbox{
We consider $d$-dimensional Riemanian manifolds which admit $d-2$ commuting
space-like Killing vector fields, orthogonal to a surface, containing two
one-parametric families of light-like curves. The condition of the  Ricci
tensor to  be zero gives Ernst equations for the metric. We write explicitly a
family of local solutions of this equations corresponding to arbitrary initial
data  on two characteristics in terms of a series. These metrics  describe
scattering of 2 gravitational   waves, and thus we expect they  are very
interesting.

Ernst equations can be written as  equations of motion for some  2D 
Lagrangian, which governs fluctuations of the metric, constant in the Killing
directions. This  Lagrangian  looks essentially as a 2D 
chiral field model, and thus is possibly treatable in the quantum case by
standart methods. It is conceivable that it may  describe
physics of some  specially arranged scattering experiment,  thus giving an 
insight for 4D gravity, not treatable by standart quantum field theory
methods.

The renormalization flow for  our Lagrangian is different from the flow for the
unitary chiral field model, the difference is essentially due to the fact that
here the field is taking values in a non-compact space of symmetric matrices.
We  investigate the model and derive the  renormalized action in
one loop.
}

\end{abstract}
\newpage

\section{Introduction}
Exact solutions of Einstein equations were of considerable intererest for a
long time and have important applications, say for cosmology; they also serve
as backgrounds for  quantum field theories, and they give an
approximate solution for metrics in the target space for conformally
invariant string theories.

There were many interesting approaches to get  solutions of Einstein
equations, most of them using symmetries, or Killing vectors. One of the most
interesting discovery was the existence of an infinite-dimensional group of
transformations of
metrics, mapping a solution of Einstein equation to another solution
(Geroch group).

If we have 2 commuting Killing vector fields in 4 dimensions, we can use them 
to reduce the problem
to two dimensions. With some additional conditions on the Killing
vectors satisfied, the free Einstein equations can be written in a nice form
of Ernst equations, which admit a Lax pair representation as a compatibility
condition of two auxiliary linear equations with a spectral parameter
\cite{ernst}, \cite{zakh},
\cite{thanasisernst}.

In the paper, we will write explicitly the local solutions of Ernst equation.
We have effectively all such solutions, since we can satisfy arbitrary
initial-boundary data. The solution is in term of a series. It is a very
interesting   problem to understand what are the properties of these
solutions and
what kind of singularities they develop.

If we use our
Killing vectors to reduce the problem to two dimensions, we get an interesting
Lagrangian in 2D, describing fluctuations of the metric,
which are constant in
the direction of our Killing vector fields. This
Lagrangian is quite close toa  principal chiral field model, but for  fields
with values
in symmetric
matrices.  If we take this
Lagrangian, but for fields with values in the $SU(n)$ group instead,
(with $n=2$ for
reduction from 4 dimensions), we get just the principal chiral field, which
admit an exact solution
in the quantum case. We expect that this Lagrangian is treatable by standart
methods, and that, unlike the unitary chiral field, it's not asymptotically
free. We  discuss the renormalization flow for this Lagrangian in one loop in
the  section 3.

\section{Solutions of the Einstein Equation in Various Dimensions,
Depending from 2 Light-Cone Parameters.}
We consider Riemanian manifolds of dimension $d=3,4,\ldots 6,\ldots
10,11,12,\ldots$  such that \hfill\break
1)they admit $(d-2)$ space-like commuting
Killing vectors fields, orthogonal to a $(u,v)$ surface, which contains two
one-parametric families of light-like curves. Thus it is posssible to introduce
coordinates $x=(u, v ,x^2, x^3, \ldots
x^d))$,
such that in that coordinates
\bge
\bga
d s^2 \equiv {\tilde{g}}_{\mu \nu} (x) d x^\mu d x^\nu =g_{ab} (u,v)d
x^a d x^b -2 f(u,v) du dv,
\\
a,b = 2, 3,\ldots d.
\ena
\ene

\vspace{16pt}

\noindent 2) The
metric ${\tilde{g}}_{\mu \nu} (x)$  is Ricci flat, 
\bge
R_{\mu \nu}=0,
\ene
where $R_{\mu \nu}$ is the Ricci tensor.

Let us introduce a function $\varphi(u,v)$ as follows:
\bge
f(u,v)= G^{-\frac{1}{2}} (u,v) \exp (2 \varphi (u,v)),
\ene
where
$$
G(u,v) = det( g_{ab}(u,v)).
$$.

The Ricci flatness condition yeilds the following equations:
\bge
\begin{array}{r}
R_{mn}=0, \quad m,n = 2,3\ldots d \Rightarrow \\
\qquad {\left(G^{\frac{1}{2}} g_u g^{-1}\right)}_v +
{\left(G^{\frac{1}{2}} g_v g^{-1}\right)}_u =0 \label{mn}
\ena
\ene

Taking Trace in (~\ref{mn}), we obtain
\bge
{\left(G^{\frac{1}{2}}\right)}_{uv}=0
\label{wave}
\ene

The other nontrivial equations are
\bge
\begin{array}{r}
R_{uu}=0 \Rightarrow
\\
\varphi_u = -\dst\frac{1}{2} \dst\frac{{\left(G^{\frac{1}{2}}\right)}_{uu}}
{{\left(G^{\frac{1}{2}}\right)}_{u}} -\dst\frac{1}{8}
\dst\frac{G^{\frac{1}{2}}}
{{\left(G^{\frac{1}{2}}\right)}_{u}} Tr \left(g_u g^{-1} g_u g^{-1} \right)
\\[20pt]
R_{vv}=0 \Rightarrow
\\
\varphi_v = -\dst\frac{1}{2} \dst\frac{{\left(G^{\frac{1}{2}}\right)}_{vv}}
{{\left(G^{\frac{1}{2}}\right)}_{v}} -\dst\frac{1}{8}
\dst\frac{G^{\frac{1}{2}}}
{{\left(G^{\frac{1}{2}}\right)}_{v}} Tr\left(g_v g^{-1} g_v g^{-1}
\right),\label{uuvv}
\ena
\ene
and
\bge
\begin{array}{r}
R_{uv}=0 \Rightarrow
\\[15pt]
\varphi_{uv}=\dst\frac{1}{8}
Tr \left(g_u g^{-1} g_v g^{-1} \right)\label{uv}.
\ena
\ene

Using (~\ref{mn}), (~\ref{wave}), we can show that the equations (~\ref{uuvv})
for $\varphi$ are compatible,
$$
\varphi_{uv}= \varphi_{vu}=\dst\frac{1}{8}
Tr \left(g_u g^{-1} g_v g^{-1} \right),
$$
which coinsides
with (~\ref{uv}). If we are given $g$, which is a solution
of (~\ref{mn}), then (~\ref{uuvv}) is a compatible system of linear
equations for $\varphi$, which can be easily integrated. Therefore, we
need to solve only equations (~\ref{mn}).

{}From (~\ref{wave}) it follows that
\bge
a(u,v) \equiv G^{\frac{1}{2}}(u,v) = y(v) - x(u)
\ene
Let us choose the local coordinates $x$ and $y$ instead of $u,v$.
In the new coordinates, the equations (~\ref{mn}) become
\bge
\bga
{((y-x) g_x g^{-1})}_y +  {((y-x) g_y g^{-1})}_x = 0,\quad -1\leq x< y \leq 1
\\[10pt]
g(-1,y) \ \ \mbox{and} \quad g(x,1) \mbox{are given.} \label{errb}
\ena
\ene
\subsection{Linear Ernst Equation and Abel Transform.}
Let $g(x,y)$ be a diagonal matrix,
$$
g(x,y) = \pmatrix{e^{\ga^{(1)}(x,y)}&0\cr 0& e^{\ga^{(2)}(x,y)}\cr}
$$
In that case, the equation (~\ref{errb}) reduces to a pair of linear equations
of the form
\bge
\bga
2(y-x) {\ga}_{xy} + {\ga}_{x}-{\ga}_{y}=0,\quad i=1,2.\label{linearernst}
\ena
\ene
\newtheorem{rns}{Proposition}
\begin{rns}
The solution of (~\ref{linearernst}) is given by
\bge
\bga
\ga(x,y) = \dst\frac{1}{\pi i
}\dst\int_{-1}^{x}
\dst\frac{{(\la-1)}^{\frac{1}{2}}}
{{(\la-x)}^{\frac{1}{2}}
{(\la-y)}^{\frac{1}{2}}}\hat{\ga_1}(\la) d\la - 
\dst\frac{1}{\pi i }
\dst\int_{y}^{1}
\dst\frac{{(\la+1)}^{\frac{1}{2}}}{{(\la-x)}^{\frac{1}{2}}
{(\la-y)}^{\frac{1}{2}}}\hat{\ga_2}(\la) d\la
\\[10pt]
\hat{\ga_1}(\la)=\dst\int_{-1}^{\la}\dst\frac{
\ga_{x}(x,1)}{{(\la-x)}^{\frac{1}{2}}} dx,\quad
\hat{\ga_2}(\la)=\dst\int_{\la}^{1}\dst\frac{
\ga_{y}(-1,y)}{{(\la-y)}^{\frac{1}{2}}} dy
\ena
\ene
\end{rns}
\newtheorem{remer}{Remark}
\begin{remer}
Let us take $y=1$. Then we will obtain a transformation
\bge
\bga
\ga(x) = \dst\frac{1}{\pi i
}\dst\int_{-1}^{x}\dst\frac{1}{{(\la-x)}^{\frac{1}{2}}}\hat{\ga_1}(\la) d\la\\
\hat{\ga_1}(\la)=\dst\int_{-1}^{\la}\dst\frac{
\ga_{x}(x,1)}{{(\la-x)}^{\frac{1}{2}}} dx
\ena
\ene
Such transformation is an Abel transform.
\end{remer}

\subsection{The Nonlinear Ernst Equation}
We need $g(x,y)$ to be a real symmetric matrix, with the determinant equal to
${(x-y)}^2$. We first ignore these conditions and solve the equation for a
general $n\times n$ complex matrix; then we will show how to satisfy the
additional conditions.

\noindent {\bf Theorem.} {\it Consider the Ernst equation
\bge
{((y-x) g_x g^{-1})}_y +  {((y-x) g_y g^{-1})}_x = 0,\label{err}
\ene
where $g(x,y)$ is a complex $n\times n $ matrix.

The following series,
\bge
\bga
g(x,y)= \left( I - 2 \dst\int \dst\frac{J(\la_1)}{p(\la_1)} d\la_1 +\right.
\\[14pt]
\left. +2\dst\sum_{n=2}^{\infty}\dst\int \pmatrix{0,& I\cr} {Z^-}_{n,n-1}
{Z^-}_{n-1,n-2}\ldots {Z^-}_{2,1}\pmatrix{-\dst\frac{F(\la_1)}{p(\la_1)} \cr
-\dst\frac{J(\la_1)}{p(\la_1)}\cr}\right.
\\
\left. \hspace{50pt} d\la_1 d\la_2 \ldots d\la_n \right.\Biggr)
g_0,\label{g,ser}
\ena
\ene
where
$$
\bga
p(\la):={(\la-x)}^{\frac{1}{2}} {(\la-y)}^{\frac{1}{2}}
\\[16pt]
{Z^\mp}_{nm}\equiv Z^\mp (\la_n, \la_m)=
\pmatrix{\dst\frac{F(\la_n)}{\la_n -\la_m}\dst\frac{p(\la_m)}{p(\la_n)}, &
\dst\frac{J(\la_n)}{\la_n -\la_m} \mp \dst\frac{F(\la_n)}{p(\la_n)}\cr
\dst\frac{J(\la_n)}{\la_n -\la_m} \dst\frac{p(\la_m)}{p(\la_n)}, &
\dst\frac{F(\la_n)}{\la_n -\la_m} \mp \dst\frac{J(\la_n)}{p(\la_n)}\cr}
\\[20pt]
F(\la)\quad \mbox{is an arbitrary} \ \ C^\infty \quad  n\times n \mbox{ \
matrix}\ ,
\\[20pt]
J(\la)=\left\{\bga \mbox{an arbitrary} \ C^\infty\quad  n\times n \mbox{\
matrix}\\[18pt]
\Theta(\la +1) \Theta (1-\la)\Theta(y-x) \Big(J_1(\la)\Theta(x-\la)+
J_2(\la)\Theta(\la-y )\Big),
\\
\mbox{where} \ J_1(\la),\  J_2(\la)\  \mbox{\ are   arbitrary\ }\  C^\infty \
\mbox{matrices, such that} \ J_1(-1)=J_2(1)=0
\ena\right.
\\[30pt]
I\  \mbox{\ is an identity matrix},
\\[12pt]
g_0\  \mbox{\ is an arbitrary constant matrix},
\ena
$$
is a solution of the Ernst equation (~\ref{err}).
}

The proof is based on the following lemmas 1 and 2.

\newtheorem{lm}{Lemma.}
\vspace{12mm}

\begin{lm}

Let
\bge
\nu(x,y,\la_0)=2\left(I +\dst\sum_{n=1}^{\infty}
\pmatrix{0,& I\cr}\dst\int {Z^-}_{n,n-1} {Z^-}_{n-1,n-2}\ldots
{Z^-}_{1,0}\pmatrix{1 \cr 1 \cr} d\la_n d\la_{n-1}\ldots d\la_1
\right)\label{nuer,ser},
\ene
then
\bge
\bga
\frac{1}{2} g^{-1}(x,y) \nu(x,y,\la_0) =
\\[10pt]
={g_0}^{-1}\left(I +\dst\sum_{n=1}^{\infty}
\dst\int \pmatrix{0,& I\cr} {Z^+}_{n,n-1} {Z^+}_{n-1,n-2}\ldots
{Z^+}_{1,0}\pmatrix{1 \cr 1 \cr} d\la_n d\la_{n-1}\ldots d\la_1 \right),
\ena
\ene
where $g(x,y)$ is the series (~\ref{g,ser}).
\end{lm}
\vspace{8mm}

The proof is by induction.
\vspace{25pt}

\begin{lm}
Let  $g(x,y)$ and $\nu(x,y,\la_0)$ be given by the series (~\ref{g,ser}) and
(~\ref{nuer,ser}), respectively.
Then
\bge
\bga
\dst\frac{\partial}{\partial x}\nu(x,y,\la_0) = \frac{1}{2}\left(1+
{\left(\dst\frac{\la_0-y}{\la_0- x}\right)}^{\frac{1}{2}}\right) g_x g^{-1}
\nu(x,y,\la_0),
\\[10pt]
\dst\frac{\partial}{\partial y}\nu(x,y,\la_0) = \frac{1}{2}\left(1+
{\left(\dst\frac{\la_0-x}{\la_0- y}\right)}^{\frac{1}{2}}\right) g_y g^{-1}
\nu(x,y,\la_0)
\label{laxer,ser}
\ena
\ene
as formal series.
\end{lm}

\noindent{\em Proof.}

We define the generalized function $x^\la,$ and ${x_+}^\la$, $\la \neq -1, -2,
\ldots$ as in \cite{1a}. The integrals like
$$
\dst\int_{0}^{\infty} f(x) x^{-\frac{3}{2}} dx
$$
should be  understood as the analytic continuation of the
integral
$$
\dst\int_{0}^{\infty} f(x) x^{\la} dx.
$$
The generalized functions $x^\la$ have the properties
\bge
\bga
\frac{d}{dx}(x^\la)= \la x^{\la-1},\quad \la\neq 0
\\
x x^{\la} = x^{\la+1}. \label{genfun}
\ena
\ene
All the kernels in (~\ref{g,ser}) and (~\ref{nuer,ser}) and their derivatives
are well-defined
as generalized functions. Using the properties (~\ref{genfun}) of these
generalized functions, we will work
with the kernels as with usual functions.

Since the formulas (~\ref{laxer,ser}) are symmetric in $x,y$, it is enough to
prove the first
equation in  (~\ref{laxer,ser}) only.

Using Lemma 1 to compute the product $\frac{1}{2} g^{-1}(x,y) \nu(x,y,\la_0)$,
the identity we need to prove in degree $n$ in $Z$ reads:
\bge
\bga
\dst\frac{\partial}{\partial x}\pmatrix{0,& I\cr}
{Z^-}_{n,n-1} {Z^-}_{n-1,n-2}\ldots {Z^-}_{1,0}\pmatrix{1 \cr 1 \cr} -
\\
- \left(1+
{\left(\dst\frac{\la_0-y}{\la_0- x}\right)}^{\frac{1}{2}}\right)\dst\sum_{m}
\dst\frac{\partial}{\partial x}\left( \pmatrix{0,& I\cr} {Z^-}_{n,n-1}
{Z^-}_{n-1,n-2}\ldots {Z^-}_{m+1,m}\pmatrix{-\dst\frac{F_m}{p_m} \cr
-\dst\frac{J_m}{p_m}\cr}\right)\cdot
\\
\qquad \pmatrix{0,& I\cr}{Z^+}_{m-1,m-2}
{Z^+}_{m-2,m-3}\ldots {Z^+}_{1,0}\pmatrix{1 \cr 1 \cr} =0\label{id1}
\ena
\ene
where
$$
p_k\equiv p(\la_k)\equiv {(\la-x)}^{\frac{1}{2}} {(\la-y)}^{\frac{1}{2}}
$$, and the block-diagonal matrices $Z$ are defined in ( ~\ref{g,ser}).

Applying the Leibnits rule and using the identities
$$
\bga
\dst\frac{\partial}{\partial x} {Z^\pm}_{k,k-1} = -\dst\frac{1}{2}
\pmatrix{\dst\frac{F_k}{p_k (\la_k -x)} \cr \dst\frac{J_k}{p_k (\la_k-x)}\cr}
\pmatrix{{\left(\dst\frac{\la_{k-1}-y}{\la_{k-1}- x}\right)}^{\frac{1}{2}}
,&\pm 1 \cr}
\\[15pt]
\dst\frac{\partial}{\partial x}\pmatrix{-\dst\frac{F_k}{p_k} \cr
-\dst\frac{J_k}{p_k}\cr} =
-\dst\frac{1}{2}
\pmatrix{\dst\frac{F_k}{p_k (\la_k -x)} \cr \dst\frac{J_k}{p_k (\la_k-x)}\cr}
\\[15pt]
\pmatrix{\dst\frac{F_m}{p_m} \cr \dst\frac{J_m}{p_m}\cr}\pmatrix{0,& I\cr}=
\dst\frac{1}{2} \left({Z^+}_{m,m-1} - {Z^-}_{m,m-1}\right),
\ena
$$
the left-hand side of (~\ref{id1}) can be rewritten as
$$
\bga
\dst\sum_{k} (0,I) \ {Z^-}_{n,n-1} {Z^-}_{n-1,n-2}\ldots {Z^-}_{k+1,k}\cdot W_k
\\
\mbox{where}
\\[20pt]
W_k = \left( \dst\frac{\partial}{\partial x} {Z^-}_{k,k-1}\right)
{Z^-}_{k-1,k-2}\ldots {Z^-}_{1,0} \pmatrix{1 \cr 1 \cr} -
\\[10pt]
- (1 +
{(\frac{\la_0-y}{\la_0- x})}^{\frac{1}{2}}) \dst\sum_{m}  \left(
\dst\frac{\partial}{\partial x} {Z^-}_{k,k-1}\right)\cdot
\\
\hspace{15pt} {Z^-}_{k-1,k-2}\ldots {Z^-}_{m+1,m}
\pmatrix{-\dst\frac{F_m}{p_m} \cr -\dst\frac{J_m}{p_m}\cr}
\pmatrix{0,& I\cr}
{Z^+}_{m-1,m-2} {Z^+}_{m-2,m-3}\ldots {Z^+}_{1,0}
\pmatrix{1 \cr 1 \cr} +
\\[10pt]
+ \dst\frac{1}{2} (1 +
{(\frac{\la_0-y}{\la_0- x})}^{\frac{1}{2}})
\pmatrix{\dst\frac{F_k}{p_k (\la_k -x)} \cr \dst\frac{J_k}{p_k (\la_k-x)}\cr}
{Z^+}_{k-1,k-2} {Z^+}_{k-2,k-3}\ldots {Z^+}_{1,0}
\pmatrix{1 \cr 1 \cr}=
\\[20pt]
=
 \dst\frac{1}{2}
\pmatrix{\dst\frac{F_k}{p_k (\la_k -x)} \cr \dst\frac{J_k}{p_k (\la_k-x)}\cr}
\ \ \Bigl({W^-}_{k-1} -{W^+}_{k-1} \Bigr),
\ena
$$
where
$$
{W^\pm}_{k}= (1\pm
{(\frac{\la_0-y}{\la_0- x})}^{\frac{1}{2}})
\pmatrix{\mp {\left(\dst\frac{\la_{k}-y}{\la_{k}- x}\right)}^{\frac{1}{2}} ,& 1
\cr}
{Z^\pm}_{k,k-1} {Z^\pm}_{k-1,k-2} {Z^\pm}_{1,0} \pmatrix{1 \cr 1 \cr}.
$$
Define also
$$
{V^\pm}_{k}= (1\pm
{(\frac{\la_0-y}{\la_0- x})}^{\frac{1}{2}})
\pmatrix{1,&\mp{\left(\dst\frac{\la_{k-1}-y}{\la_{k-1}-
x}\right)}^{\frac{1}{2}}\cr}
{Z^\pm}_{k,k-1} {Z^\pm}_{k-1,k-2} {Z^\pm}_{1,0} \pmatrix{1 \cr 1 \cr}
$$

We will prove  by induction that
\bge
\bga
\\
{W^+}_{k} ={W^-}_{k}
\\
{V^+}_{k} ={V^-}_{k}.\label{id2}
\ena
\ene

For $k=1$, the identities (~\ref{id2}) can be checked easily.
Using the identities,
\bge
\bga
\pmatrix{\mp {\left(\dst\frac{\la_{k}-y}{\la_{k}- x}\right)}^{\frac{1}{2}} ,& 1
\cr}
{Z^\pm}_{k,k-1}=
\\
= \dst\frac{F_k}{(\la_k -\la_{k-1})}\dst\frac{p_{k-1}}{p_k}
\pmatrix{1,&\mp {\left(\dst\frac{\la_{k-1}-y}{\la_{k-1}-
x}\right)}^{\frac{1}{2}}  \cr} +
\dst\frac{J_k}{(\la_k -\la_{k-1})}\dst\frac{(\la_{k-1}-x)}{(\la_{k}-x)}
\pmatrix{\mp {\left(\dst\frac{\la_{k-1}-y}{\la_{k-1}- x}\right)}^{\frac{1}{2}}
,& 1 \cr};
\\[20pt]
\pmatrix{1 ,& \mp {\left(\dst\frac{\la_{k}-y}{\la_{k}- x}\right)}^{\frac{1}{2}}
\cr}
{Z^\pm}_{k,k-1}=
\\
\dst\frac{J_k}{(\la_k -\la_{k-1})}\dst\frac{p_{k-1}}{p_k}
\pmatrix{1,&\mp {\left(\dst\frac{\la_{k-1}-y}{\la_{k-1}-
x}\right)}^{\frac{1}{2}}  \cr}
+
\dst\frac{F_k}{(\la_k -\la_{k-1})}\dst\frac{(\la_{k-1}-x)}{(\la_{k}-x)}
\pmatrix{\mp {\left(\dst\frac{\la_{k-1}-y}{\la_{k-1}- x}\right)}^{\frac{1}{2}}
,& 1 \cr}
\ena
\ene
we can reduce an expression with  product of $k$ matrices $Z$ in (~\ref{id2})
to a similar expression with product of only $k-1$ matrices $Z$, which proves
the lemma by induction assumption.

\vspace{20pt}
The proof of the Theorem follows from Lemmas 1,2 and form the fact that the
compatibility condition
for equations (~\ref{laxer,ser}) gives the Ernst equation (~\ref{err}). Since
we have shown that with $g(x,y)$ given by (~\ref{g,ser}) both equations of
(~\ref{laxer,ser}) are satisfied, the compatibility conditions is fulfilled
automatically, and thus $g(x,y)$ solves the equation (~\ref{err}).

\begin{lm}
Let
\bge
\bga
F(\la)=0\\
J(\la):=\Theta(\la +1) \Theta (1-\la)\Theta(y-x) \Big(J_1(\la)\Theta(x-\la)+
J_2(\la)\Theta(\la-y )\Big),\\
g_0 = I
\ena
\ene
where $\Theta$ is the Heaviside functon. Then $g(x,y)$ given by (~\ref{g,ser})
is real. If $J_1$ and $J_2$ are symmetric matrices, then $g(x,y)$  is a
symmetric matrix.
\end{lm}
{\em Proof}
With our choise of the branch of the square root, the reality condition is
obvious. The fact that we get a symmetric matrix can be proved by induction.
\vspace{20pt}

We can ensure that the determinat of the matrix $g(x,y)$ is ${(x-y)}^2$ by
specifying $g(x,1)$ and $g(-1,y)$ and finding recursively $J_1(\la)$ and
$J_2(\la)$, using the defining relation (~\ref{g,ser}) and the inversion
formula for
the Abel transform, discussed in the linear case.

\section{Quantum version.} The equations (~\ref{mn}), (~\ref{uv}) are the
Euler-Lagrange equations for
the action
\bge
S=\frac{1}{\al}\dst\int {(det g)}^{\frac{1}{2}} \left( \varphi_{uv} +  Tr
\left( g_u g^{-1} g_v g^{-1}\right)\right) du\wedge dv,
\ene
This action, up to a total derivative, is just the Einstein action,
specialized for metrics which are constant along the integral lines of the
Killing vector fields and written for metrics in the   gauge
$$
d s^2 = g_{ab} (u,v)d
x^a d x^b -2 {(det(g))}^{-\frac{1}{2}} \exp (\frac{1}{4} \varphi (u,v)) du dv.
$$
This Lagrangian may  actually describe some specially arranged scattering
experiment, probably the  fluctuations in the background of  two colliding
gravitational plane waves, where the measurements are averaged in the normal to
the collision plane directions by the detector, so only the constant in those
directions mode survive.

In any event, this Lagrangian gives an interesting two dimensional model, worth
studying in the quantum case, especially since firstly it has something to do
with gravity, and secondly it is  not that infinitely far from being treatable
by conventional methods, as the 4 dimensional gravity is. It looks essentially
as the  chiral field Lagrangian, with the field $g$ taking values not in the
unitary group, but in symmetric positive-definite matrices, namely,  in
$2\times 2$ matrices for reduction from 4 dimensions. (If we would take $g$
with values in the $SU(n)$ group instead, the determinant of $g$ is
automatically 1, which trivially solves the wave equation, and our Lagrangian
becomes  just the principal chiral field Lagrangian of $SU(n)$; such model was
solved exactly in the quantum case,  \cite{polw}.)  Our Lagrangian  has a
global $SL(2,\Bbb R)$ symmetry,
\bge
g \rightarrow h^T g h, \quad h\in SL(2,\Bbb R) \label{sl2r},
\ene
corresponding to basis changes in the Killing directions.
The power counting for this Lagrangian works in the same way as for  the
chiral field, or a $\sigma$ model,  so the issue of renormalizability is quite
clear. However, the renormalization flow is  different, compared to the unitary
chiral field, and, essentially, is more like a renormalization flow for the
$SL(2,\Bbb R)$ chiral field (such model, with the WZW term added, was solved
recently in \cite{zml^2}; it is curious however that for symmetric $2\times
2$matrices the WZW term $Tr\left( dg g^{-1}{\wedge}^{*3}\right)$ is identically
zero).
The rough argument here is to use the $\sigma$-model language
and to say that the $\beta$-function is proportional to the
Ricci tensor, which is no longer positive-definite, and therefore
it's not an asymptotically free case, as it was for a unitary chiral field.

When reducing from 4 dimensions, the Killing directions  should be thought of
as, say,  compactified  on a torus, then the integral over those
non-interesting coordinates gives simply the area of the torus, and we get a
Planck constant in front which is of order ${(l_{planck}/l_\perp)}^2$, where
$l_\perp$ is
a typical length in directions normal to the collision plane; this is really  a
pretty small number, thus we expect that the loop expansion should be
reliable, and we will consider the quantum corrections to the Lagrangian in one
loop only. Our treatment is similar to that for a $\sigma$ model, \cite{gwdzk},
\cite{frdn}, \cite{gsw}. It could be very helpful to use the  technique 
developed for the chiral field model, like in \cite{pol}, for the case of
non-unitary or non-compact group. 

Unlike the case of the $\sigma$- model or principal chiral field, there is a
natural additive structure for   symmetric matrices, so  we just expand
around, say, one of the classical solutions $g, \varphi$ found in the
beginning of the paper,
$$
\bga
\tilde{g}(u,v)=g (u,v) + h(u,v)\\
\tilde{\varphi(u,v)}=\varphi (u,v) + \phi (u,v),
\ena
$$
and keep only the quadratic terms in fluctuations, to get
\bge
S= \frac{1}{\al} S_0 + \langle \nabla_u \xi, \nabla_v \xi \rangle +
\xi R \xi,
\ene
where $S_0$ is, accidently, zero on equations of motion, and 
$
\xi ={\pmatrix{h_{11},&h_{22},&h_{12},&\phi \cr}}^T ,
$
and $\langle \ \ \rangle$  and  $R$ can be  expressed
in terms of  the components of the background fields $g , \varphi$, as follows
\bge
\bga
\langle \xi \eta  \rangle\equiv \xi^T m \eta,
\\
m={(det(g))}^{-\frac{3}{2}}\pmatrix{
{(g_{22})}^2 & {(g_{12})}^2 & - 2 g_{22} g_{12}& -\frac{1}{4} g_{22} det(g)\cr
{(g_{12})}^2 & {(g_{11})}^2 & - 2 g_{11} g_{12}& -\frac{1}{4} g_{11} det(g)\cr
- 2 g_{22} g_{12}& - 2 g_{11} g_{12} & 2 (g_{11} g_{22}+ {(g_{12})}^2 ) &
\frac{1}{2} g_{12} det{g}\cr
-\frac{1}{4} g_{22} det(g) & -\frac{1}{4} g_{11} det(g) & \frac{1}{2} g_{12}
det{g} & 0 \cr}
\ena
\ene
Although $m$ does not transform as a metric under linear transformations of
$\xi$, we use it to  define an $SL(2,\Bbb R)$-invariant scalar product,
(Minkowski signature), with
$SL(2, \Bbb R)$ acting as (~\ref{sl2r}).
We can find an orthogonal reper for this scalar product $e_{(\al)}$,
$\al=1,2,3,4$, of the form
$$
(e_{(1)}, e_{(2)}, e_{(3)}, e_{(4)}= \pmatrix{* & * & * & *\cr
0 & * & * & *\cr 0 & 0 & * & * \cr 0 & 0 & 0 & *}
$$,
such that, for $g$ positive-definite, and
\bge
\langle e_{(\al)}  e_{(\beta)} \rangle = \eta\equiv diag (1, 1 ,1,-1);
\label{reper}
\ene
$\nabla=\partial + a $ is defined as follows:
\bge
\bga
\nabla_u h = {\partial}_u h - h g^{-1} g_u - g_u g^{-1} h +
\frac{1}{2}Tr(g^{-1} h) g_u -\frac{1}{4 \langle e_{(4)} e_{(4)}\rangle}
{(det(g))}^{\frac{1}{2}}  Tr(g^{-1}h g^{-1} g_u) e_{(4)} \\
\nabla_u \chi = {\partial}_u \chi \label{nabla}
\ena
\ene

\bge
\bga
\xi\  R\  \xi = {det(g)}^{\frac{1}{2}}  \Biggr( \ \Biggl.
\\[14pt]
\left( -\frac{1}{8} {\left(Tr (g^{-1} h )\right)}^2 -\frac{1}{4} Tr(g^{-1} h
g^{-1} h )\right) Tr( g_u g^{-1} g_v g^{-1})+ \\
+  \left( \frac{1}{8} {\left(Tr (g^{-1} h )\right)}^2 -\frac{1}{4} Tr(g^{-1} h
g^{-1} h )\right)\varphi_{uv} +\\
+Tr(g^{-1} h)Tr(g_\mu g^{-1} g^\mu g^{-1} h g^{-1}) - Tr(g_u g^{-1} h g^{-1}
g_v g^{-1} h g^{-1})-\\
 -\frac{1}{16} {(det(g))}^{\frac{1}{2}} Tr(g_u g^{-1} h g^{-1}) Tr( g_v g^{-1}
h g^{-1}) \Biggr. \Biggl)
\ena
\ene

Substituting  $\xi = \chi^\al e_{(\al)}$ in the Lagrangian we get
\bge
S= \partial \chi \partial \chi + A \chi A \chi + 2\partial\chi A\chi +
\chi (e\ R\ e) \chi,
\ene
where
\bge
A= \tilde{e}(\partial +a) e,
\ene
with $a$ defined in (~\ref{nabla}), and $\tilde{e}$ is dual to the reper $e$,
(~\ref{reper}), $${\tilde{e}}^{(\al)}_i {e}_{(\al)}^j =\delta_{i}^{j}.$$
We now switch to the Eucledean version, so $u,\ v$ should be now $z, \bar{z}$,
and the metric $\eta$ is Eucledean, and use the dimensional regularization to
compute Feynman diagrams.

There are two lograrithmically divergent Feynmann diagrams with the 
background $A$
fields, contributing to the effective action, the bubble and the loop with
2 $\chi$ propagators and 2 $\partial\chi A\chi$ vertices. The divergent
contribution of the antisymmetric
in internal indices part of $A$ coming from these two diagrams cancels one
another, due to the fact that
\bge
\bga
\int \frac{d^d k}{k^2 +\varepsilon^2}\sim \Gamma(1-\frac{d}{2})
\varepsilon^{d-2}= \frac{1}{2-d} + \mbox{regular} \\
\int \frac{k_\mu k_\nu}{(k^2 +\varepsilon^2)({(k-q)}^2 + \varepsilon^2)} d^d
k\sim \frac{1}{2} \delta_{\mu \nu}\Gamma(1-\frac{d}{2})
\int_{0}^{1}dt 
{(\varepsilon^2 + t(1-t) q^2)}^{\frac{d-2}{2}} + \mbox{regular}=\\
=\frac{1}{2} \delta_{\mu \nu} \frac{1}{2-d} + \mbox{regular},
\ena
\ene
and the factors of $2$ and minus signs comes just right for those diagrmas
to cancel one
another; we have put in a mass to cut off in the infrared, but nothing depend
from it, and there many ways to fix the infrared problems, 
so we are not concerned with that.
For the symmetric in internal indices part  of $A$, the loop diagram is not
divergent, and the bubble diagram is, so there is a total contribution of the
$A$ terms to the effective action. 
Adding it to  the contribution of the bubble diagram with
$(e \ R \ e)$, we get the following correction to the effective action in one
loop:
\bge
\frac{\mu^{d-2}}{2-d}\Bigl(\frac{1}{2} (A_{\al_1 \al_2} A_{\al_1 \al_2}
-A_{\al_1 \al_2} A_{\al_2 \al_1}) + e_{(\al)}^{i} \ R_{ij} e_{(\al)}^{j} \Bigr)
\ene
Those correction terms  have two 'world-sheet' derivatives in them, so they can
be thought of as a correction to what is 
the metric in the $\sigma$-model language,
and therefore, the meaning of the above computation is in that it shows what 
is the 
replacement in our case for the well-known fact about $\sigma$ models, that 
the $\beta$
function is proportional to the Ricci tensor. In the $\sigma$ model, the 
renormalizations can be organized to correct the metric and to produce 
the dilaton and tahion potentials; however, I have not found a convenient way
to keep track of the corrections for this model. 
\section{Discussion}
We have constructed explicitly local solutions of free Einstein equations,
in the case when there are $d-2$ commuting Killing vector fields. We have, 
in fact, all local solutions, since we can satisfy the apropriate initial
data.  
It will be very interesting to understand the global properties of 
these solutions, and
what kind of singularities they develop. To achieve that, we firstly need
a better control over the convergence of the series. For  integrable
equations like the nonlinear Schrodinger, we can prove that the perturbation
series is convergent, and we are working on the convergence for the Ernst.
Secondly, our solutions are written in particular coordinates, which make
the computation easy. However, to understand the singularities, we need to
formulate everything invariantly.

We proposed a two dimensional model, describing quantum fluctuations of
the metric, constant along the Killing directions. This model is interesting
just by itself, since it is close to the chiral field model, and 
thus apparently
treatable, but has a different renormalization flow in the ultraviolet, and,
most probably, is not asymptotically free. We have studied  the
renormalization in one loop. This  model is definitely not an easy one, and
more have to be done to understand where it actually flows. 

There is another reason of why the model is interesting, since it might
describe some specially arranged scattering experiment, when only those
two dimension we keep matters. If so, this will give an insight for
gravity in 4 dimension, which is not treatable by standart field theory 
methods.

Our 2 dimensinal model can be also  looked at  as some kind of string theory,
but with the world-sheet having quite concrete interpretation in terms of the
target space geometry.

It will be interesting to have the supersymmetric version of this model.

\end{document}